\documentclass[aps,prb,twocolumn,showpacs]{revtex4}

\usepackage{amsmath}
\usepackage{graphicx}
\usepackage{setspace}
\usepackage{bm}
\usepackage{dcolumn}
\usepackage{ulem}

\begin{document}

\title{Flux exchange in inhomogeneous type-II superconductors}

\author{Rongchao Ma} 
\affiliation{Department of Physics, University of Alberta, Edmonton, Alberta, Canada}
\email{marongchao@yahoo.com}

\date{\today}

\begin{abstract}
The vortex hopping motion in a type-II superconductor determines the current-carrying ability and consequently the application fields of the superconductor. However, it is not clear how the vortices hop between the different pinning regions in the superconductor. Here we proposed that there should be magnetic \textit{flux exchange} between two contacting pinning regions. A system of differential equations was constructed to describe the flux exchange phenomenon. The qualitative analysis methods were used. The approximate numerical solutions and approximate analytical solutions of the system were obtained. The results show that the flux exchange reduces the internal field in a weak pinning region, but increases the internal field in a strong pinning region. Moreover, the flux exchange phenomenon is strongly influenced by the superconductor's geometrical size.
\end{abstract}


\maketitle

\section{Introduction}

The vortices in a type-II superconductor are generally pinned down by pinning centers.\cite{Yeshurun,Tinkham,Blatter}
But there is a probability for a vortex to spontaneously hop between the adjacent pinning centers due to the vortex's thermal fluctuation.\cite{Yeshurun,Tinkham,Blatter,Tonomura,Sanders,Fisher,Anderson}
The hopping motion can be described by the Arrhenius equation, which shows that the hopping frequency is strongly related to the vortex's activation energy.\cite{Anderson}
The hopping motion causes the magnetic flux (or current) in the superconductor to reduce with increasing time, which is usually referred as flux relaxation. This phenomenon exhibits various time evolution behaviors because of the complicated field (or current) dependence of the vortex activation energy.\cite{Anderson,Beasley,Larkin,Feigel'man1,Feigel'man2,Zeldov1,Zeldov2,Ma1,Ma2}
Due to its dissipation, the hopping motion determines the superconductor's current carrying ability and consequently its application fields.\cite{Palau,Koelle,Yeshurun,Tinkham} 
Thus, a large number of experimental and theoretical research works \cite{Silhanek1,Silhanek2,Rosenstein,Basov,Yeshurun,Blatter,Palau,Koelle,Sonier,Maniv,Altshuler} have been carried out to study the vortex hopping motion.

In the present work, we wish to study the mutual vortex hopping motion between the contacting pinning regions with different pinning ability, which we shall call \textit{flux exchange}. More specifically, a real superconductor may be divided into smaller regions whose sizes are larger than the superconductor's penetration depth. Because of the different pinning ability of the smaller regions,\cite{Schrieffer} the vortices close to the interface between two contacting pinning regions will hop from one region into another region \cite{Kalisky1,Kalisky2,Gurevich}, and vice versa. In other words, the two contacting pinning regions exchange vortices (flux). Because of the different pinning ability, the vortex hopping motions in each smaller region need to be described by different equations. This indicates that a flux exchange process is determined by the parameters of both contacting pinning regions. It can not be simply described by the conventional flux relaxation theory. Thus, we need to consider the \textit{flux exchange} seriously and construct new mathematical equations to describe it.

However, the flux exchange is usually ignored in the magnetization (or internal field) measurements carried out over an entire sample.\cite{Yeshurun,Tinkham,Shen,Kim} To study the flux exchange phenomenon, we may consider the fact that the vortices in a weak pinning region have a higher energy and higher hopping frequency, but the vortices in a strong pinning region have a lower energy and lower hopping frequency. Consequently, there should be more vortices hop from the weak pinning region into the strong pinning region in a unit time interval. This will result in a nonzero vortex migration from the weak pinning region to the strong pinning region, which changes the vortices quantity in each pinning region. Thus, it is possible to study the flux exchange phenomenon by investigating its influences on the flux quantity in each pinning region.

In this work, the physical principle of the flux exchange phenomenon was analyzed using the concepts of ``average internal fields''. First, we constructed a system of differential equations to describe the flux exchange phenomenon by investigating its influence on the flux quantity in each pinning region. Next, we calculated the approximate numerical solutions and approximate analytical solutions of the system to view the time evolution behavior of the flux exchange. Finally, we discussed the flux exchange between the bulk pinning region and surface pinning region, the space outside the superconductor, respectively.

\section{Principle of flux exchange}

As mentioned before, the flux exchange in this work means the mutual vortex hopping motion between the contacting pinning regions. This motion occurs in a thin layer close to the interfaces between the contacting pinning regions. It indicates that one can study the flux exchange by watching the pinning regions' edge and counting how many vortices hopped into and hopped out of the pinning region. On the other hand, a vortex must first go to a pinning region's edge and then can hop out from this pinning region. Thus, one can also obtain the information about the vortex motion in the pinning region's center by studying the flux exchange at the pinning region's edge.

\subsection{Prelude}

Let us now consider an inhomogeneous superconductor which includes many small regions with different pinning ability (see Figure 1). The materials in each small region are uniform. Below its critical temperature $T_c$, each pinning region has a maximum internal field $B_m$ (or melting field $H_m$) due to its limited pinning ability. Above $B_m$, the vortices cannot be pinned down and exhibit fluidity. Below $B_m$, the vortices are pinned down and can only move through thermal fluctuation by hopping between adjacent pinning centers with a probability.\cite{Tinkham}

\begin{figure}[htb]
\label{figure1}
\begin{center}
\includegraphics[height=0.25\textwidth]{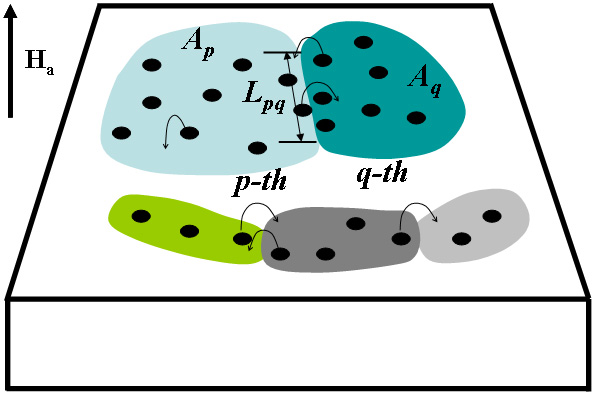}
\caption{(Color online) Schematic diagram of magnetic flux exchange between the $p$-th pinning region and $q$-th pinning region of a type-II superconductor. The black disks represent vortices. The arrows represent the vortices' hopping directions. $A_p$ is the cross area perpendicular to the magnetic field in the $p$-th pinning region. $A_q$ is the cross area perpendicular to the magnetic field in the $q$-th pinning region. $L_{pq}$ is the line length on the interface perpendicular to the magnetic field.
}
\end{center}
\end{figure}

We shall focus on the vortex hopping motion (below $B_m$) in this section, but discuss the vortex flowing motion (above $B_m$) at the end of this paper. The vortex hopping motion can be described by the Arrhenius equation $\nu = \nu_0 \mathrm{e}^{-U/kT}$, where $\nu$ is hopping frequency, $\nu_0$ is attempt frequency, $U$ is vortex activation energy, $k$ is Boltzmann constant, and $T$ is temperature. Multiplying $\nu$ by an average hopping distance $s$, we obtain the vortex velocity \cite{Beasley,Vinokur}
\begin{equation}
\label{VV}
v = s \nu_0 \mathrm{e}^{-U/kT}. 
\end{equation}

Let $B_{im}$ be the maximum internal field of the $i$-th pinning region. Apply a uniform field, $B_a \leq min(B_{1m},\cdots,B_{Nm})$, to the superconductor at room temperature and then cool the superconductor down to a temperature $T$ below its transition temperature $T_c$. A uniform starting field is then realized in the sample. When the applied magnetic field is turned off, the vortices in various pinning regions are then in a non-equilibrium state. This is because the vortex migration rate depends on $vB$, where $v$ is vortex velocity and $B$ is vortex density. Eq.(\ref{VV}) shows that $v$ varies in different pinning regions, which means that a uniform vortex density cannot maintain an equilibrium state between different pinning regions. Thus, the vortices start to migrate from a weak pinning region to a strong pinning region. As the vortices migration goes on, a vortex gradient could be formed over the interface. According to Maxwell's equation, a persistent current may flow around the pinning regions' edge to maintain the vortex density gradient.

Any current in a superconductor exerts a Lorentz force on the vortices, such as the persistent current circulating in a ring shape superconductor (or a cylindrical superconductor) for maintaining the flux trapped in the superconductor's hole, or the injected current flowing in a superconducting wire. The Lorentz force can reduce the vortex activation energy $U$ and increase the vortex hopping rate. Thus, $U$ is usually expressed as a decreasing function of the current density $j$, i.e., $U(j)$.\cite{Anderson} On the other hand, each current has a corresponding magnetic field according to the Biot-Savart law. The Lorentz force on a vortex is then equivalent to the expelling force from other vortices. In addition to the vortices generated by the currents, there are also some trapped vortices in the superconductor (say, a uniformly distributed field) that is unrelated to any bulk currents. Thus, $U$ can also be expressed as a decreasing function of the internal field $B$, i.e., $U(B)$.\cite{Beasley} We shall use this field dependent expression in the following discussions. One can have a clearer picture of this problem by referring to the magnetic phase diagram ($H-T$ diagram).\cite{Blatter} In this diagram, all the physical properties and dynamical behavior (relaxation, phase transition) of a vortex system are represented by two parameters, the field $H$ and temperature $T$.

To avoid mathematical complexity and focus on physical analysis, we shall use the concept ``average internal field'' \cite{Houghton} in each pinning region although there may be a flux distribution in the pinning region. This is possible because the vortex hopping motion can be considered globally and therefore the local vortex density distribution can be ignored.\cite{Geshkenbein} Let us now define the average internal field in the $i$-th pinning region as,
\begin{equation}
\label{FieldB}
B_i = \frac{\Phi_i}{A_i} = \frac{1}{A_i} \int \vec{B}_i \cdot d\vec{A}_i,
\end{equation} 
where $A_i$ is the area of the surface perpendicular to the magnetic field and $\Phi_i=\int \vec{B}_i \cdot d\vec{A}_i$ is the magnetic flux in this region. With definition Eq.(\ref{FieldB}), we shall further apply a cut off to the vortex activation energies on the interface where the flux exchange occurs.

\subsection{Differential equations of flux exchange}

To describe the flux exchange phenomenon with mathematical language, we need to construct a system of differential equations. This can be achieved by considering the rate of change of the magnetic flux in each pinning region, which is proportional to vortex hopping frequency \cite{Anderson,Geshkenbein,Ma1}, or vortex velocity (see Eq.(\ref{VV})). To this end, we shall first consider a flux exchange process that only involves two pinning regions: the $p$-th pinning regions and $q$-th pinning regions (whose sizes are larger than the superconductor's penetration depth, see Figure. 1). Later, we shall generalize it into a flux exchange process that involves $N$ different pinning regions (see system (\ref{sde})).

In the $p$-th pinning region, the magnetic flux variation in a time interval $\Delta t$ includes two parts:
(a) Due to flux exchange, the vortices (close to the interface) in the $p$-th pinning region hop into the $q$-th pinning region \cite{Beasley,Vinokur} by $ -\Delta \Phi_{pq} = - {}^1/_2 B_p L_{pq} v_p \Delta t$, where $B_p$ is the average internal field (vortex density) in the $p$-th pinning region, $L_{pq}$ is the line length on the interface perpendicular to field, $v_p$ is the vortex velocity in the $p$-th pinning region (see Eq.(\ref{VV})). The number ${}^1/_2$ accounts for the random hopping motion in both directions.
(b) Due to flux exchange, the vortices in the $q$-th pinning region hop back to the $p$-th pinning region by $ \Delta \Phi_{qp} = {}^1/_2 B_q L_{qp} v_q \Delta t$, where $B_q$ is the average internal field in the $q$-th pinning region, $L_{qp}=L_{pq}$, and $v_q$ is the vortex velocity in the $q$-th pinning region.

The total change of magnetic flux in the $p$-th pinning region is now $ \Delta \Phi_p = -\Delta \Phi_{pq} + \Delta \Phi_{qp}$. Letting $A_p$ be the cross area perpendicular to the magnetic field in the $p$-th pinning region and dividing both sides by $A_p \Delta t$, we obtain a differential equation for the average internal field in the $p$-th pinning region, that is,
\begin{equation}
\label{Epq0}
\frac{\mathrm d B_p}{\mathrm d t} = - \epsilon_{pq} B_{pq}, 
\end{equation}
where $\epsilon_{pq}= {}^1/_2 L_{pq} s \nu_0/A_p$ and
\begin{equation}
\label{Bpq}
B_{pq}= B_p \mathrm{e}^{-U_p/kT} - B_q \mathrm{e}^{-U_q/kT}
\end{equation}
is a function related the flux exchange (Eq.(\ref{VV}) is used).

In the $q$-th pinning region, similarly, the total change of magnetic flux is $ \Delta \Phi_q = - \Delta \Phi_{qp} + \Delta \Phi_{pq}$. Letting $A_q$ be the cross area perpendicular to the magnetic field in the $q$-th pinning region and dividing both sides by $A_q \Delta t$, we obtain a differential equation for the average internal field in the $q$-th pinning region, that is,
\begin{equation}
\label{Eqp0}
\frac{\mathrm d B_q}{\mathrm d t} = - \epsilon_{qp} B_{qp}, 
\end{equation}
where $\epsilon_{qp}= {}^1/_2 L_{qp} s \nu_0/A_q$ and $B_{qp}=-B_{pq}$.

Eq.(\ref{Epq0}) and Eq.(\ref{Eqp0}) constitute a nonlinear system of differential equations that describes the flux exchange between the $p$-th pinning region and $q$-th pinning region. From mathematics we know that there is no general method available for calculating the exact solutions of a nonlinear system.\cite{Cushing} Thus, we will study the nonlinear system using qualitative analysis methods, approximate numerical methods, and approximate analysis methods.

\section{Solutions of flux exchange}

To study the nonlinear system Eq.(\ref{Epq0}), Eq.(\ref{Eqp0}), we need the initial values $B_p(0)$ and $B_q(0)$. Furthermore, we still need the detailed expressions of the activation energies $U_p$ and $U_q$. A number of activation energies were proposed in early studies.\cite{Anderson,Beasley,Larkin,Feigel'man1,Feigel'man2,Zeldov1,Zeldov2,Ma1,Ma2} In principle, we can choose any one of these activation energies for both $U_p$ and $U_q$. If the number of activation energies is $N$, then the number of systems of differential equations is $N^2$. This is a big number and it is impossible to exhaust all the possible combinations in the present paper. For simplicity, we shall calculate the solutions only using the following two combinations: \\

\noindent (a). Linear activation energy \cite{Anderson,Ma1}: 
\begin{equation}
\label{LinearAE}
\left\{
\begin{aligned}
&U_p(B_p)=U_{p0}(1-B_p/B_{pm}), \\
&U_q(B_q)=U_{q0}(1-B_q/B_{qm}).
\end{aligned}
\right.
\end{equation}

\noindent (b). Logarithmic activation energy \cite{Zeldov1,Zeldov2}: 
\begin{equation}
\label{LogAE}
\left\{
\begin{aligned}
&U_p(B_p) = U_{p0} ln(B_{pm}/B_p), \\
&U_q(B_q) = U_{q0} ln(B_{qm}/B_q).
\end{aligned}
\right.
\end{equation}

Using Eq.(\ref{LinearAE}) and Eq.(\ref{LogAE}), we see that the right-hand sides of system Eq.(\ref{Epq0}), Eq.(\ref{Eqp0}) do not explicitly depend on the variable $t$. Thus, system Eq.(\ref{LinearAE}) and Eq.(\ref{LogAE}) is an autonomous system.

\subsection{Existence and uniqueness}

From physics we know that the solution of system Eq.(\ref{Epq0}), Eq.(\ref{Eqp0}) do exist and is unique on the intervals $B_p \in [0,B_{pm}]$, $B_q \in [0,B_{qm}]$. However, we still need to prove it in a pure mathematical point of view.

The fundamental existence and uniqueness theorem of differential system states that \cite{Cushing}, if all the partial derivatives, 
\begin{equation}
\label{PartialDeri0}
\frac{\partial B'_p}{\partial B_p}, ~\frac{\partial B'_p}{\partial B_q}, ~\frac{\partial B'_q}{\partial B_p}, ~\frac{\partial B'_q}{\partial B_q},
\end{equation}
are continuous on some intervals, then the initial value problem of system Eq.(\ref{Epq0}), Eq.(\ref{Eqp0}) has a unique solution on these intervals.

Using Eq.(\ref{Epq0}), Eq.(\ref{Eqp0}), we have
\begin{equation}
\label{PartialDeri}
\left\{
\begin{aligned}
&\frac{\partial B'_i}{\partial B_i} = - \epsilon_{ij} \left( 1 - \frac{  B_i}{kT} \frac{\partial U_i}{\partial B_i} \right) e^{-U_i/kT}, ~i \neq j, \\
&\frac{\partial B'_i}{\partial B_j} = \epsilon_{ij} \left(1-\frac{B_j}{kT} \frac{\partial U_j}{\partial B_j} \right) e^{-U_j/kT},
\end{aligned}
\right.
\end{equation}
where $i=p,q$ and $j=p,q$.

\noindent 1. Linear activation energy

Using Eq.(\ref{LinearAE}), we have 
\[ \frac{\partial U_i}{\partial B_i}= - \frac{U_{i0}}{B_{im}}. \]

Substituting into Eq.(\ref{PartialDeri}), one can see that the partial derivatives in Eq.(\ref{PartialDeri0}) are continuous on $(-\infty,\infty)$. Thus, the initial value problem of system Eq.(\ref{Epq0}), Eq.(\ref{Eqp0}) with the linear activation energy Eq.(\ref{LinearAE}) has a unique solution on $B_p \in (-\infty,\infty)$, $B_q \in (-\infty,\infty)$.

\noindent 2. Logarithmic activation energy

Using Eq.(\ref{LogAE}), we have 
\[ \frac{\partial U_i}{\partial B_i}= - \frac{U_{i0}}{B_i}. \]

Substituting into Eq.(\ref{PartialDeri}), one can see that the partial derivatives in Eq.(\ref{PartialDeri0}) does not exist at the point $B_p=0$, or $B_q=0$. Thus, the initial value problem of system Eq.(\ref{Epq0}), Eq.(\ref{Eqp0}) with the logarithmic activation energy Eq.(\ref{LogAE}) has a unique solution in real number field except the zero points $B_p=0$, or $B_q=0$.

\subsection{Phase plane portraits}

Generally, it is difficult to draw the phase portrait of a nonlinear system.\cite{Cushing} Fortunately, the $B_p(t)$ and $B_q(t)$ in system Eq.(\ref{Epq0}), Eq.(\ref{Eqp0}) has a simple linear relationship. This makes it very easy to draw the phase portrait. Let us now prove it.

Canceling out $B_{pq}$ (or -$B_{qp}$) from Eq.(\ref{Epq0}), Eq.(\ref{Eqp0}), we have 
\begin{equation}
\label{SysMerg}
\mathrm d B_p = - \gamma \mathrm d B_q.
\end{equation}
where $\gamma = \epsilon_{pq}/\epsilon_{qp}$.

Integrating both sides of Eq.(\ref{SysMerg}) with respect to the time $t$, we have
\begin{equation}
\label{BpBq}
B_p(t) = - \gamma \left[ B_q(t) - B_q(0) \right] + B_p(0),
\end{equation}
where $B_p(0)$, and $B_q(0)$ are the initial value of $B_p(t)$ and $B_q(t)$, respectively.

This shows that $B_p(t)$ and $B_q(t)$ have a linear relationship. Therefore, the time evolution behaviors of the internal fields in each pinning region are similar although their activation energies are different. Using Eq.(\ref{BpBq}), one can easily draw the phase portrait of system Eq.(\ref{Epq0}), Eq.(\ref{Eqp0}), as shown in Figure 2.

\begin{figure}[htb]
\label{figure2}
\begin{center}
\includegraphics[height=0.3\textwidth]{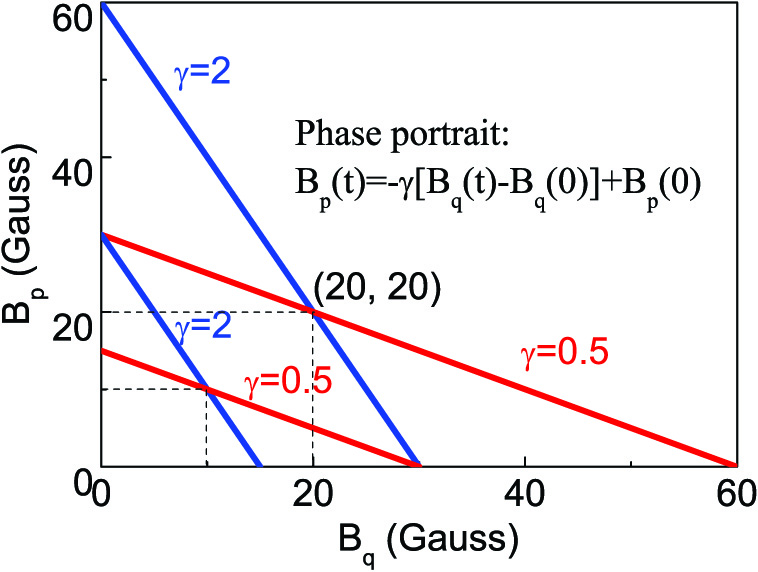}
\caption{(Color online) 
Phase portrait of system Eq.(\ref{Epq0}), Eq.(\ref{Eqp0}). The average internal fields $B_p(t)$ and $B_q(t)$ have a linear relationship, $B_p(t) = - \gamma \left[ B_q(t) - B_q(0) \right] + B_p(0)$ (see Eq.(\ref{BpBq})). The slopes were put as $\gamma$ = 0.5 and 2, respectively. The initial conditions were put as $B_p(0)=B_q(0)=20$ G and $B_p(0)=B_q(0)=10$ G, respectively.
}
\end{center}
\end{figure}

\subsection{Equilibrium point}

An equilibrium point (or critical point) \cite{Cushing} is a point where $B'_p(t)=0$, $B'_q(t)=0$. From Eq.(\ref{Epq0}) and Eq.(\ref{Eqp0}), we see that the net flux exchange vanishes at the equilibrium point, i.e., $B_{pq}=-B_{qp}=0$. Using Eq.(\ref{Bpq}), we have
\begin{equation}
\label{ep}
kT ln \left(\frac{B_p}{B_q}\right) = U_p(B_p) - U_q(B_q).
\end{equation}

\noindent 1. Linear activation energies

Substituting Eq.(\ref{LinearAE}) into Eq.(\ref{ep}), we have
\begin{equation}
\label{LinearEP}
ln \left(\frac{B_p}{B_q}\right) + \left( \tilde{\alpha}_p B_p -  \tilde{\alpha}_q B_q \right) = \left( \alpha_p - \alpha_q \right).
\end{equation}
where $\alpha_p=U_{p0}/kT$, $\tilde{\alpha}_p=\alpha_p/B_{pm}$, $\alpha_q=U_{q0}/kT$, and $\tilde{\alpha}_q=\alpha_q/B_{qm}$.

The equilibrium point $B_p$, $B_q$ is determined by Eq.(\ref{LinearEP}) and Eq.(\ref{BpBq}).

\noindent 2. Logarithmic activation energies

Substituting Eq.(\ref{LogAE}) into Eq.(\ref{ep}), we have
\begin{equation}
\label{LogEP}
\frac{B_p^{\alpha_p+1}}{B_q^{\alpha_q+1}} = \frac{B_{pm}^{\alpha_p}}{B_{qm}^{\alpha_q}}.
\end{equation}

The equilibrium point $B_p$, $B_q$ is determined by Eq.(\ref{LogEP}) and Eq.(\ref{BpBq}).

\subsection{Numerical solution}

To obtain a direct view on the effects of flux exchange, we did approximate numerical calculations for the $B_p(t)$, $B_q(t)$ in Eq.(\ref{Epq0}), Eq.(\ref{Eqp0}). The activation energies were chosen from Eq.(\ref{LinearAE}) and Eq.(\ref{LogAE}). The parameters were put as follows: $B_{pm} = 20$, $B_{qm} = 100$ G, $U_{p0}/kT=15$, $U_{q0}/kT=75$, and $\epsilon_{pq}$=$\epsilon_{qp}$=0, 0.001, 0.01, 0.1, and 0.5 s$^{-1}$ respectively. The initial condition is $B_p(0)=B_q(0)=20$ G. In the numerical calculations, we noticed that the step of time $t$ must be very small. Otherwise, the results are unstable. This indicates that system Eq.(\ref{Epq0}), Eq.(\ref{Eqp0}) is a stiff system. The calculated results are shown in Figure 3.

\begin{figure}[htb]
\label{figure3}
\begin{center}
\includegraphics[height=0.6\textwidth]{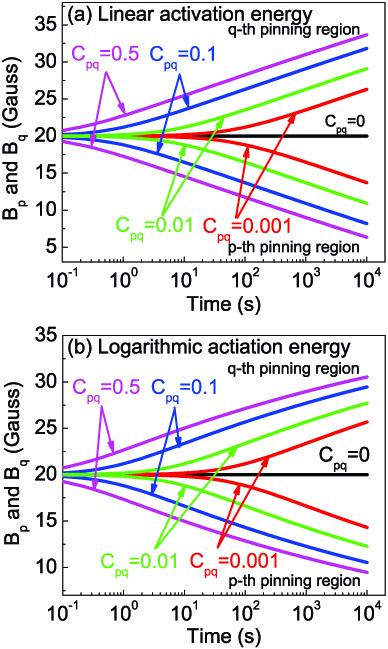}
\caption{(Color online) 
Numerical solutions of magnetic flux exchange. In the calculations, we used the following two activation energy combinations for the $p$-th and $q$-th pinning regions, respectively: 
(a) Linear activation energy $U_p(B_p)=U_{p0}(1-B_p/B_{pm})$ and $U_q(B_q)=U_{q0}(1-B_q/B_{qm})$. 
(b) Logarithmic activation energy $U_p(B_p) = U_{p0} \ln(B_{pm}/B_p)$ and $U_q(B_q) = U_{q0} \ln(B_{qm}/B_q)$.
The parameters were put as: $B_{pm} = 20$ G, $B_{qm} = 100$ G, $U_{p0}/kT=15$, $U_{q0}/kT=75$, and $\epsilon_{pq}$=$\epsilon_{qp}$ = 0, 0.001, 0.01, 0.1 and 0.5 s$^{-1}$ respectively. The initial condition is $B_p(0)=B_q(0)=20$ G.
}
\end{center}
\end{figure}

Figure 3 shows that the flux exchange reduces $B_p(t)$ (the internal field in the $p$-th pinning region), but increases $B_q(t)$ (the internal field in the $q$-th pinning region). It looks that the $q$-th pinning region attracts vortices from the $p$-th pinning region. In fact, there is no attracting force in the $q$-th pinning region. It is just because we have assumed that the vortex activation energy $U_p < U_q$, and therefore, the vortex hopping frequency $\nu_p > \nu_q$. This difference results in a ``deficit'' of flux exchange in the $p$-th pinning region, but a ``surplus'' of flux exchange in the $q$-th pinning region.

\subsection{Approximate analytical solutions}

In the following paragraphs, we shall explore the analytical solutions of Eq.(\ref{Epq0}), Eq.(\ref{Eqp0}) under the approximation of ``significant flux exchange'', i.e., $|B_{pq}|=|B_{qp}| >> 0$. This occurs in two cases: \\
\noindent (1). $B_p\mathrm{e}^{-U_p/kT} >> B_q\mathrm{e}^{-U_q/kT}$, then $B_{pq}\approx B_p\mathrm{e}^{-U_p/kT}$; \\
\noindent (2). $B_p\mathrm{e}^{-U_p/kT} << B_q\mathrm{e}^{-U_q/kT}$, then $B_{pq}\approx -B_q \mathrm{e}^{-U_q/kT}$.

Due to the symmetry between case (1) and case (2), we only need to calculate the solutions in case (1). Afterwards, we can obtain the solutions of case (2) by simply doing a commutation to
the subscripts $p \leftrightarrow q$.

Substituting $B_{pq}\approx B_p\mathrm{e}^{-U_p/kT}$ into Eq.(\ref{Epq0}), we have 
\begin{equation}
\label{Eqj1a}
\frac{\mathrm d B_p}{\mathrm d t} = - \epsilon_{pq} B_p\mathrm{e}^{-U_p/kT}. 
\end{equation}

One can obtain $B_p(t)$ by solving Eq.(\ref{Eqj1a}) and obtain $B_q(t)$ using Eq.(\ref{BpBq}). Let us now use the initial condition $B_p(0)=B_q(0)=B_a$ and start with the linear activation energy.

\subsubsection{Linear activation energy}

Substituting the linear activation energy Eq.(\ref{LinearAE}) into Eq.(\ref{Eqj1a}), we have
\begin{equation}
\label{Eqj2a}
\frac{\mathrm{e}^{-x}}{x} \mathrm d x = -\epsilon_{pq}\mathrm{e}^{-\alpha_p} \mathrm d t,
\end{equation}
where
\begin{equation}
\label{Funcx}
x=\tilde{\alpha}_p B_p,
\end{equation}
$\alpha_p=U_{p0}/kT$, $\tilde{\alpha}_p=\alpha_p/B_{pm}$.

Using the formula
\[ \int \frac{\mathrm{e}^{-x}}{x} \, \mathrm{\mathrm d } x = \ln|x| + \sum\limits_{n=1}^\infty \frac{(-x)^n}{n\cdot n !} \]
and integrating both sides of Eq.(\ref{Eqj2a}) with respect to $t$, we have
\begin{equation}
\label{FuncBi}
\ln|x| + \sum\limits_{n=1}^\infty \frac{(-x)^n}{n\cdot n !} = - \left( \epsilon_{pq}\mathrm{e}^{-\alpha_p} \right) t + C_1,
\end{equation}
where
\[ C_1 =  \ln|x(0)| + \sum\limits_{n=1}^\infty \frac{[-x(0)]^n}{n\cdot n !}  \] 
is an integrating constant determined by the initial condition $x(0)=\tilde{\alpha}_p B_a$ (or $B_p(0)=B_a$).

The internal field in the $p$-th pinning region, $B_p(t)$, is then determined by Eq.(\ref{Funcx}) and Eq.(\ref{FuncBi}). $B_q(t)$ can be obtained by substituting $B_p(t)$ into Eq.(\ref{BpBq}).

\subsubsection{Logarithmic activation energy}

Substituting the logarithmic activation energy $U_p(B_p) = U_{p0} \ln(B_{pm}/B_p)$ into Eq.(\ref{Eqj1a}), we have
\begin{equation}
\label{BpLogD}
B_p^{-(\alpha_p+1)} \mathrm dB_p = -\epsilon_{pq} B_{pm}^{-\alpha_p} \mathrm d t
\end{equation}
where $\alpha_p=U_{p0}/kT$.

Integrating both sides of Eq.(\ref{BpLogD}) with respect to $t$, we have
\begin{equation}
\label{BpLog}
B_p(t) = B_a \left( 1 + \frac{t}{\tau_p} \right)^{-1/\alpha_p}.
\end{equation}
where $\tau_p= B_{pm}^{\alpha_p}/(\alpha_p \epsilon_{pq} B_a^{\alpha_p})$.

$B_q(t)$ can be obtained by substituting Eq.(\ref{BpLog}) into Eq.(\ref{BpBq}).

\section{Generalization of flux exchange and flux relaxation}

We have studied a flux exchange process that involves two pinning regions $p$ and $q$. However, we are still not clear how the flux exchange behavior is exactly in a real superconductor that includes many pinning regions. In this section, we shall first generalize system Eq.(\ref{Epq0}), Eq.(\ref{Eqp0}). Afterwards, we shall use it to describe the flux exchange and flux relaxation over the entire superconductor.

\subsection{Generalized flux exchange equations}

The system Eq.(\ref{Epq0}), Eq.(\ref{Eqp0}) describes a flux exchange process that only involves two different pinning regions. If a flux exchange process includes $N$ different pinning regions, then the system of differential equations should be generalized into
\begin{widetext}
\begin{equation}
\label{sde}
\left\{
\begin{aligned}
\frac{\mathrm d B_1}{\mathrm d t} =& - \epsilon_{12} B_{12} - \cdots - \epsilon_{1j} B_{1j} - \cdots - \epsilon_{1N} B_{1N}, \\
\frac{\mathrm d B_2}{\mathrm d t} =& - \epsilon_{21} B_{21} - \cdots - \epsilon_{2j} B_{2j} - \cdots - \epsilon_{2N} B_{2N}, \\
\cdots\cdots \\
\frac{\mathrm d B_i}{\mathrm d t} =& - \epsilon_{i1} B_{i1} - \cdots - \epsilon_{ij} B_{ij} - \cdots - \epsilon_{iN} B_{iN}, ~(j \neq i)\\
\cdots\cdots \\
\frac{\mathrm d B_N}{\mathrm d t} =& - \epsilon_{N1} B_{N1} - \cdots - \epsilon_{Nj} B_{Nj} - \cdots - \epsilon_{N(N-1)} B_{N(N-1)},
\end{aligned}
\right.
\end{equation}
\end{widetext}
where $\epsilon_{ij}= {}^1/_2 L_{ij} s \nu_0/A_i$, $B_{ij}= B_i \mathrm{e}^{-U_i/kT} - B_j \mathrm{e}^{-U_j/kT}$, and $i=1,\cdots, N$. If the $i$-th pinning region and $j$-th pinning region are not contacted, then their interface length $L_{ij}=L_{ji}=0$. Consequently, $\epsilon_{ij}=\epsilon_{ji}=0$ and the corresponding exchange term vanishes automatically (Note that generally $\epsilon_{ij} \neq \epsilon_{ji}$ unless $A_i=A_j$). It should be emphasized that system (\ref{sde}) is only applicable for vortex hopping motion (below $B_m$), but not for vortex flowing motion (above $B_m$), where $B_m$ is the maximum internal field (see section II). Thus, the initial magnetic fields must satisfy the inequality $B_i(0) \leq B_{im}$, where $B_{im}$ is the maximum internal field in the $i$-th pinning region.

For simplicity, let us rewrite system (\ref{sde}) as:
\[ \frac{\mathrm d B_i}{\mathrm d t} = - \sum\limits_{j'=1}^N \epsilon_{ij} B_{ij}, \]
where $i,j=1,\cdots, N$. The $'$ on $j$ means $j \neq i$, i.e., the summation does not include $\epsilon_{ii}$ terms.

\subsection{Flux exchange over the entire superconductor and flux relaxation}

A superconductor usually includes many pinning regions. Among all these regions, the superconductor's surface (surface pinning region, see Figure 4) play a special role.\cite{Sun,Pautrat,Burlachkov,Beasley,Bean} In the following paragraphs, we shall use system (\ref{sde}) to describe the flux exchange between the surface pinning region and bulk pinning region, the space outside the superconductor. Afterwards, we shall discuss the flux relaxation process over the entire superconductor.

\begin{figure}[htb]
\label{figure4}
\begin{center}
\includegraphics[height=0.23\textwidth]{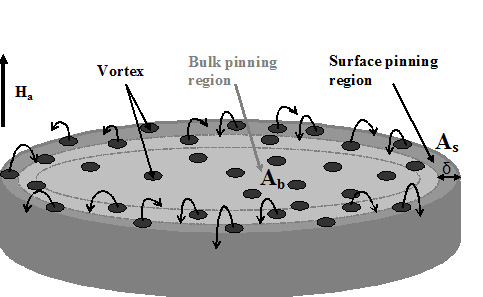}
\caption{(Color online) 
Schematic diagram of magnetic flux exchange between the surface pinning region and bulk pinning region, the space outside a type-II superconductor, respectively. The black disks represent vortices. The arrows represent the vortices' hopping directions. $A_b$ is the cross area perpendicular to the magnetic field in the bulk pinning region. $A_s$ is the cross area perpendicular to the magnetic field in the surface pinning region. $\delta$ is the thickness of the surface pinning region.
}
\end{center}
\end{figure}

\subsubsection{Flux exchange with surface pinning region}

The surface pinning region \cite{Beasley} is a thin layer close to the superconductor's surface that is parallel to the magnetic field. The thickness of this layer is approximately equal to the penetration depth. In the surface pinning region, the vortices are subjected to a strong surface pinning \cite{Sun,Pautrat,Burlachkov,Beasley,Bean} and have a lower energy.\cite{Yeshurun} In the superconductor's bulk, however, the vortices are subjected to weak random pinnings \cite{Dolz,Larkin,Feigel'man1} and have a higher energy \cite{Yeshurun}. 
Thus, the surface pinning region and bulk pinning region should exchange magnetic flux (see Figure 4). On the other hand, the surface pinning region also exchanges flux with the space outside the superconductor. This will change the magnetic flux quantity in the superconductor and have a strong impact on the vortex behavior. Thus, it is important to study the flux exchange processes that involve the surface pinning region. For simplicity, let us assume that the entire bulk pinning region is uniform and can be treated as a single `bulk pinning region'.

Let $B_{sm}$ be the maximum internal field of the surface pinning region. At the beginning, there are more vortices hopping into than hopping out from the surface pinning region, which results in $B_s(t)>B_{sm}$. At this stage, the vortices in the surface pinning region should be in a liquid state and most of the vortices hop in from the bulk pinning region will flow out from the surface pinning region directly. Thus, the vortex motion in the surface pinning region cannot be described by system (\ref{sde}). The over all magnetic flux will show a time evolution behavior similar to that in the bulk pinning region, which can be approximately described by Eq.(\ref{Eqj1a}), namely, 
\begin{equation}
\label{EqBPR}
\frac{\mathrm d B_b}{\mathrm d t} = - \epsilon_{bs} B_b \mathrm{e}^{-U_b/kT}. 
\end{equation}
The solutions of Eq.(\ref{EqBPR}) is already discussed in Eq.(\ref{FuncBi}) and Eq.(\ref{BpLog}).

The vortex density and hopping rate reduces with increasing time. In the surface pinning region, it finally reaches $B_s(t)< B_{sm}$. At this stage, all the vortices hopped out from the bulk pinning region will be first trapped in the surface pinning region, and then hop out from the surface pinning region into the space outside the superconductor. Thus, the over all magnetic flux will show a time evolution behavior determined by the average of those in the surface pinning region and bulk pinning region. This will be further discussed later.

To describe the flux exchange with the surface pinning region, we shall denote the surface pinning region with a subscript `s', the bulk pinning region with a subscript `b', and the space outside the superconductor with a subscript `e'. Referring to system (\ref{sde}), we have 
\begin{equation}
\label{Sys3}
\left\{
\begin{aligned}
&\frac{\mathrm d B_b}{\mathrm d t} = - \epsilon_{bs} B_{bs}, \\
&\frac{\mathrm d B_s}{\mathrm d t} = - \epsilon_{sb} B_{sb} - \epsilon_{se} B_{se}, \\
\end{aligned}
\right.
\end{equation}
where $ B_{bs} = -B_{sb} = B_b \mathrm{e}^{-U_b/kT} - B_s \mathrm{e}^{-U_s/kT} $, $ B_{se} = B_s \mathrm{e}^{-U_s/kT} - B_a $, and $B_a$ is the magnetic field in the space outside the superconductor.

The various coefficients can be evaluated by considering a cylindrical superconductor with a radius $R$. Apply an external magnetic field parallel to the superconductor's axis. Let $\delta$ be the thickness of the surface pinning region ($\delta$ is of the order of the superconductor's penetration depth, and $\delta<<R$). Therefore, the area of the bulk pinning region is $A_b=\pi (R - \delta)^2$, the area of the surface pinning region is $ A_s = \pi R^2 -\pi (R - \delta)^2 $, and the area of the space outside the superconductor is $A_e=\infty$. The length of the line on the interface perpendicular to field is now the perimeter of the bulk pinning region, i.e., $L_{bs}=L_{sb}=2 \pi (R - \delta)$, and $L_{se}=L_{es}=2\pi R$. The flux exchange coefficients are: $\epsilon_{bs}= s \nu_0/(R - \delta)$, $\epsilon_{sb}=(R-\delta)s\nu_0/[(2R-\delta)\delta]$, $\epsilon_{se}=Rs\nu_0/[(2R-\delta)\delta]$, and $\epsilon_{es}=0$.

\subsubsection{Activation energy}

In the bulk pinning region, there is a number of options for the activation energy $U_b(B_b)$. In low temperature superconductors, the strong pinning dominates. The linear activation energy $U_b(B_b)=U_{b0}\left(1-B_b/B_{bm}\right)$ gives a good description to the vortex behavior. In high temperature superconductors, however, the weak point pinning dominates. The logarithmic activation energy $U_b(B_b)=U_{b0}ln\left(B_{bm}/B_b\right)$ and inverse-power activation energy $U_b(B_b) = U_{b0}/\mu \left[ \left(B_{bm}/B_m\right)^\mu -1 \right]$ give better descriptions to the vortex behavior.

In the surface pinning region, let us now prove that the linear activation energy $U_s(B_s)=U_{s0}\left(1-B_s/B_{sm}\right)$ is more accurate than others. Recall that the vortex deformation and the nonlinear interaction between vortices result in the higher order (nonlinear) terms in the activation energies.\cite{Ma1} However, the vortex deformation in the surface pinning region is small because the vortices are parallel to each other. Moreover, the surface pinning force is stronger than the interacting force between vortices because the surface pinning is a strong pinning. Thus, we can safely ignore the higher order terms in $U_s(B_s)$ and assume that it obeys the linear law.

\subsubsection{Total internal field and geometric effect}

In an experiment, a measured internal field is usually the total average internal field over the entire superconductor. This field can be defined as
\begin{equation}
\label{BT}
B_t = \frac{\Phi_b+\Phi_s}{A} = (1-\alpha) B_b + \alpha B_s,
\end{equation}
where $A = A_b + A_s= \pi R^2$ is the total area of the surface perpendicular to the magnetic field. $B_b=\Phi_b/A_b$ and $B_s=\Phi_s/A_s$ are the average internal fields in the bulk pinning region and surface pinning region, respectively. The geometrical factor (area ratio) 
\[ \alpha = \frac{A_s}{A} = 1- \left( 1 - \frac{\delta}{R} \right)^2 \] 
is the weight of the contribution from the magnetic flux in the surface pinning region. Let $R \to \infty$, we have $\alpha \to 0$. Thus, Eq.(\ref{BT}) reduces to $B_t = B_b$. This means that, in a large superconductor, the contribution to the total average internal field $B_t$ from the magnetic flux in the surface pinning region $\alpha B_s$ is ignorable. This property is completely determined by geometry and is nothing related to physics.

\section{Discussion}

In this section, we shall prove that the system (Eq.(\ref{Epq0}), Eq.(\ref{Eqp0})) is equivalent to Maxwell's equations. We shall also discuss the necessary and sufficient condition for nonzero flux exchange, and the causes of vortex motion.

\subsection{Equivalent with Maxwell's equations}

We have constructed a system of differential equations, Eq.(\ref{Epq0}), Eq.(\ref{Eqp0}) (or system (\ref{sde})), to describe the flux exchange process. However, the vortices in a type-II superconductor are the quanta of magnetic field, their motions must obey Maxwell's equations \cite{Vinokur}, or continuity equation \cite{Beasley}. Thus, one may ask whether the system of flux exchange is equivalent to Maxwell's equations (or continuity equation)?

The early works \cite{Vinokur,Beasley} have shown that, using Maxwell's equations (or continuity equation), the vortex motion is governed by the following equation
\begin{equation}
\label{Maxwell}
\partial_t B = - \partial_x \left( B v \right),
\end{equation}
where $B$ is internal field (vortex density) and $v$ is vortex velocity (see Eq.(\ref{VV})). Thus, our task is to prove that Eq.(\ref{Epq0}), Eq.(\ref{Eqp0}) are equivalent to Eq.(\ref{Maxwell}).

As mentioned before, the vortex hopping motions in the smaller regions are described by different equations. It indicates that the right side of Eq.(\ref{Maxwell}) cannot be directly used to describe the flux exchange between two contacting pinning regions. To solve this problem, we may refer to the definition of one-sided derivatives in mathematics. Let us now consider the left derivative of $Bv$, that is,
\begin{equation}
\label{LeftDeri}
\partial_- \left( B v \right) = \lim_{h \to 0^-} \frac{B(x+h) v(x+h) - B(x) v(x)}{h}.
\end{equation}

Let $x+h$ be a point in the $p$-th pinning region and $x$ be a point in the $q$-th pinning region. Therefore, Eq.(\ref{Maxwell}) can be rewritten as
\begin{equation}
\label{Exch}
\left( \frac{\mathrm d B_p}{\mathrm d t} \right)_{exchange} = - C_p \left(B_p v_p - B_q v_q\right),
\end{equation}
where $C_p$ is an unknown constant.

Using Eq.(\ref{VV}), one can see that Eq.(\ref{Epq0}) and Eq.(\ref{Exch}) are the same equation by putting $C_p= {}^1/_2 L_{pq}/A_p$. This shows that Eq.(\ref{Epq0}) is equivalent to Maxwell's equation Eq.(\ref{Maxwell}). Similarly, one can also prove that Eq.(\ref{Eqp0}) is equivalent to Eq.(\ref{Maxwell}).

\subsection{Remarks on flux exchange and flux relaxation}

We have shown that the flux exchange equations (system Eq.(\ref{Epq0}), Eq.(\ref{Eqp0})) is equivalent to Maxwell's equation Eq.(\ref{Maxwell}). If we let the pinning region's size reduce to $0$, then the discrete system Eq.(\ref{Epq0}), Eq.(\ref{Eqp0}) reduces to the continuous equation Eq.(\ref{Maxwell}). In this sense, one can even used the idea of ``flux exchange'' to analysis the vortex motion at a point. More specifically, to study the vortex motion at the point, one can choose a small region that includes the point, and then consider the quantity of the vortices hopped into and hopped out from the small region. This indicates that any flux creep process (or flux relaxation process) can be regarded as a flux exchange process. Therefore, the concept of ``flux exchange'' is not just an idle variant of ``flux creep''. The former is more general and has richer physics than the later.

From above discussion, we see that the necessary and sufficient condition for a nonzero flux exchange between two contacting pinning regions is $B_{pq}= B_p \mathrm{e}^{-U_p/kT} - B_q \mathrm{e}^{-U_q/kT} \neq 0$ (see Eq.(\ref{Bpq})). In the continuous form, the necessary and sufficient condition for a nontrivial flux creep process is $\partial_x \left( B v \right) \neq 0$. This disproves the necessary and sufficient condition, $\partial_x B \neq 0$ (or $B$ has a nonzero gradient), assumed in some early literatures.

Eq.(\ref{Epq0}), Eq.(\ref{Eqp0}) shows that two contacting pinning regions exchange magnetic flux even in an equilibrium state although the net exchanged flux is zero. The vortex hopping motion is still strong in the equilibrium state due to the reduced vortex activation energy. Thus, $B_{pq}\neq 0$ is only the cause of a nonzero exchanged flux quantity (or $\partial_x \left( B v \right) \neq 0$ is the cause of a nontrivial flux creep process), but not a cause of vortex motion. The causes of vortex motion are the vortex thermal fluctuation and limited vortex activation energy.

\section{Conclusion}

In a type-II superconductor, the different pinning regions exchange magnetic flux (vortices). This phenomenon causes the vortices in a weak pinning region to migrate to a strong pinning region. Thus, it influences the flux quantities in each pinning region. The flux exchange phenomenon also occurs between the surface pinning region and bulk pinning region, the space outside the superconductor, respectively. But the vortices hop from the surface pinning region into the bulk pinning region has little influence on the average internal field in the bulk pinning region. In the calculation of total average internal field, the contribution from the surface pinning region is significant in a small superconductor, but is ignorable in a large superconductor. Further studies should be carried out on the solution of the generalized system of flux exchange, i.e., system (\ref{sde}).


\begin{thebibliography}{99}


\bibitem{Yeshurun}
Y. Yeshurun, A. P. Malozemoff, A. Shaulov, Rev. Mod. Phys. \textbf{68}, 911 (1996).

\bibitem{Tinkham} 
Michael Tinkham, \textit{Introduction to Superconductivity} (McGraw-Hill, New York, 1996).

\bibitem{Blatter}
G. Blatter, M. V. Feigel'man, V. B. Geshkenbein, A. I. Larkin, and V. M. Vinokur, Rev. Mod. Phys. \textbf{66}, 1125 (1994).

\bibitem{Tonomura}
A. Tonomura, H. Kasai, O. Kamimura, T. Matsuda, K. Harada, J. Shimoyama, K. Kishio and K. Kitazawa, Nature \textbf{397}, 308 (1999).

\bibitem{Sanders}
S. C. Sanders, J. Sok, D. K. Finnemore, and Qiang Li, Phys. Rev. B \textbf{47}, 8996 (1993).

\bibitem{Fisher} 
Daniel S. Fisher, Matthew P. A. Fisher, and David A. Huse, Phys. Rev. B \textbf{43}, 130 (1991).

\bibitem{Anderson}
P. W. Anderson, Phys. Rev. Lett. \textbf{9}, 309 (1962).

\bibitem{Beasley}
M. R. Beasley, R. Labusch, and W. W. Webb, Phys. Rev. \textbf{181}, 682 (1969).

\bibitem{Larkin}
A. I. Larkin and Yu. N. Ovchinnikov, J. low Temp. Phys. \textbf{34}, 409 (1979).

\bibitem{Feigel'man1}
M. V. Feigel'man, V. B. Geshkenbein, A. I. Larkin, and V. M. Vinokur, Phys. Rev. Lett. \textbf{63}, 2303 (1989).

\bibitem{Feigel'man2}
M. V. Feigel'man, V. B. Geshkenbein, and V. M. Vinokur, Phys. Rev. B \textbf{43}, 6263 (1991).

\bibitem{Zeldov1}
E. Zeldov, N. M. Amer, G. Koren, A. Gupta, R. J. Gambino, and M. W. McElfresh, Phys. Rev. Lett. \textbf{62}, 3093 (1989).

\bibitem{Zeldov2}
E. Zeldov, N. M. Amer, G. Koren, A. Gupta, M. W. McElfresh, and R. J. Gambino, Appl. Phys. Lett. \textbf{56}, 680 (1990).

\bibitem{Ma1}
Rongchao Ma, J. Appl. Phys. \textbf{108}, 053907 (2010).

\bibitem{Ma2}
Rongchao Ma, J. Appl. Phys. \textbf{110}, 063911 (2011).

\bibitem{Palau}
A. Palau, R. Dinner, J. H. Durrell, and M. G. Blamire, Phys. Rev. Lett. \textbf{101}, 097002 (2008).

\bibitem{Koelle}
D. Koelle, R. Kleiner, F. Ludwig, E. Dantsker, and John Clarke, Rev. Mod. Phys. \textbf{71}, 631 (1999).

\bibitem{Silhanek1}
A. V. Silhanek, J. Gutierrez, R. B. G. Kramer, G. W. Ataklti, J. Van de Vondel, V. V. Moshchalkov, and A. Sanchez, Phys. Rev. B \textbf{83}, 024509 (2011).

\bibitem{Silhanek2}
A. V. Silhanek, W. Gillijns, M. V. Milosevic, A. Volodin, V. V. Moshchalkov, and F. M. Peeters, Phys. Rev. B \textbf{76}, 100502 (2007)

\bibitem{Rosenstein}
Baruch Rosenstein and Dingping Li, Rev. Mod. Phys. \textbf{82}, 109 (2010).

\bibitem{Basov}
D. N. Basov and T. Timusk, Rev. Mod. Phys. \textbf{77}, 721 (2005).

\bibitem{Altshuler}
E. Altshuler and T. H. Johansen, Rev. Mod. Phys. \textbf{76}, 471 (2004).

\bibitem{Maniv}
Tsofar Maniv, Vladimir Zhuravlev, Israel Vagner, and Peter Wyder, Rev. Mod. Phys. \textbf{73}, 867 (2001).

\bibitem{Sonier}
Jeff E. Sonier, Jess H. Brewer, and Robert F. Kiefl, Rev. Mod. Phys. \textbf{72}, 769 (2000).

\bibitem{Schrieffer}
J. Robert Schrieffer and James S. Brooks, \textit{Handbook of High -Temperature Superconductivity: Theory and Experiment} (Springer, 2007).

\bibitem{Kalisky1}
B. Kalisky, J.R. Kirtley, J.G. Analytis, J.-H. Chu, A. Vailionis, I.R. Fisher, and K.A. Moler, Phys. Rev. B \textbf{83}, 064511 (2011).

\bibitem{Kalisky2}
B. Kalisky, J. R. Kirtley, E. A. Nowadnick, R. B. Dinner, E. Zeldov, Ariando, S. Wenderich, H. Hilgenkamp, D. M. Feldmann, and K. A. Moler, Appl. Phys. Lett. \textbf{94}, 202504 (2009). 

\bibitem{Gurevich}
A. Gurevich and V. M. Vinokur, Phys. Rev. Lett. \textbf{83}, 3037 (1999).

\bibitem{Shen}
Bing Shen, Peng Cheng, Zhaosheng Wang, Lei Fang, Cong Ren, Lei Shan, and Hai-Hu Wen, Phys. Rev. B \textbf{81}, 014503 (2010).

\bibitem{Kim}
Mun-Seog Kim, C. U. Jung, Min-Seok Park, S. Y. Lee, Kijoon H. P. Kim, W. N. Kang, and Sung-Ik Lee, Phys. Rev. B \textbf{64}, 012511 (2001).

\bibitem{Vinokur}
V. M. Vinokur, M. V. Feigel'man and V. B. Geshkenbein, Phys. Rev. Lett. \textbf{67}, 915 (1991).

\bibitem{Houghton}
A. Houghton, R. A. Pelcovits, and A. Sudbo, Phys. Rev. B \textbf{40}, 6763 (1989).

\bibitem{Geshkenbein}
V. B. Geshkenbein, and A. I. Larkin, Sov. Phys. JETP \textbf{68}, 639 (1989).

\bibitem{Cushing}
J. M. Cushing, \textit{Differential Equations: An Applied Approach} (Pearson-Prentice Hall, Upper Saddle river, New Jersey, 2004).

\bibitem{Pautrat}
Alain Pautrat, Christophe Goupil, Charles Simon, Norbert Lutke-Entrup, Bernard Placais, Patrice Mathieu, Yvan Simon, Alexander Rykov and Setsuko Tajima, Phys. Rev. B \textbf{63}, 054503 (2001).

\bibitem{Sun}
Yang Ren Sun, J. R. Thompson, H. R. Kerchner, D. K. Christen, M. Paranthaman and J. Brynestad, Phys. Rev. B \textbf{50}, 3330 (1994). 

\bibitem{Burlachkov}
L. Burlachkov, Phys. Rev. B  \textbf{47}, 8056 (1993).

\bibitem{Bean}
C. P. Bean, J. D. Livingston, Phys. Rev. Lett. \textbf{12}, 14 (1964).

\bibitem{Dolz}
M. I. Dolz, A. B. Kolton, and H. Pastoriza, Phys. Rev. B \textbf{81}, 092502 (2010).



\end{thebibliography}
\end{document}